\documentclass[a4paper,11pt]{article}
\usepackage{pos}

\newcommand{\OmegaGW}{\Omega_{\mathrm{GW}}}

\let\oldsqrt\sqrt
\def\sqrt{\mathpalette\DHLhksqrt}
\def\DHLhksqrt#1#2{%
\setbox0=\hbox{$#1\oldsqrt{#2\,}$}\dimen0=\ht0
\advance\dimen0-0.2\ht0
\setbox2=\hbox{\vrule height\ht0 depth -\dimen0}%
{\box0\lower0.4pt\box2}}

\newcommand{\sss}[1]{{\scriptscriptstyle{#1}}}
\newcommand{\boldmathsymbol}[1]{{\ensuremath{\boldsymbol{#1}}}}

\newcommand{\uPl}{\mathrm{Pl}}

\newcommand{\usssPl}{\sss{\uPl}}

\newcommand{\calH}{\mathcal{H}}

\newcommand{\Mp}{M_\usssPl}

\newcommand{\beq}{\begin{equation}}
\newcommand{\eeq}{\end{equation}}
\newcommand{\bea}{\begin{equation}\begin{aligned}}
\newcommand{\eea}{\end{aligned}\end{equation}}

\newlength{\wsingfig}
\newlength{\wdblefig}
\newlength{\wquadfig}
\newlength{\wtriplefig}
\newcommand{\Eq}[1]{Eq.~(\ref{#1})}

\newcommand{\Fig}[1]{Fig.~{\ref{#1}}}

\newcommand{\Sec}[1]{Sec.~\ref{#1}}

\title{Gravitational wave signatures of non-singular matter bouncing cosmology in NANOGrav and beyond}

\author*[a,b,c]{Theodoros Papanikolaou}

\affiliation[a]{Scuola Superiore Meridionale, Largo San Marcellino 10, 80138 Napoli, Italy.}

\affiliation[b]{Istituto Nazionale di Fisica Nucleare (INFN), Sezione di Napoli, Via Cinthia 21, 80126 Napoli, Italy.}

\affiliation[c]{National Observatory of Athens, Lofos Nymfon, 11852 Athens, Greece.}

\emailAdd{t.papanikolaou@ssmeridionale.it}

\abstract{Non-singular matter bouncing cosmological setups are of particular interest since apart from adressing the initial singularity problem they can give rise as well to a nearly scale-invariant curvature power spectrum on scales $k<10^{4}\mathrm{Mpc}^{-1}$, favored by Cosmic Microwave Background (CMB) experiments. Interestingly enough, one can find that within such non-singular bouncing cosmological setups, curvature perturbations grow on super-horizon scales during the matter contracting phase. In this work, we account for the evolution of cosmological perturbations during the transition to the Hot Big Bang expanding Universe, finding at the end naturally enhanced curvature perturbations on very small scales at horizon-crossing time during the expanding phase. These enhanced cosmological perturbations can induce  gravitational waves (GWs) due to second order gravitational interactions and collapse as well to form primordial black holes (PBHs), with the latter acting as one of the most viable dark matter candidates. Remarkably, we find an induced GW background with a universal infrared (IR) frequency scaling of $f^2$, in excellent agreement with the recently released $\mathrm{nHz}$ GW data by the NANOGrav collaboration, being potentially detectable as well by other GW observatories such as LISA and ET, depending on the values of the bouncing cosmological parameters at hand.

}

\FullConference{Proceedings of the Corfu Summer Institute 2024 "School and Workshops on Elementary Particle Physics and Gravity" (CORFU2024)\
12 - 26 May, and 25 August - 27 September, 2024\\
Corfu, Greece\\}

\begin{document}
\maketitle

\section{Introduction}
According to the recent release of the 15-year pulsar timing array (PTA) gravitational-wave data by the NANOGrav Collaboration, there is strong evidence in favor of the existence of a gravitational-wave (GW) background at the $\mathrm{nHz}$ frequency range~\cite{NANOGrav:2023gor}, having initiating during the past two years an enormous research activity focusing in the interpretation of this low-frequency GW background. In particular, among many works in the topic, there have been proposed possible explanations of the PTA $\mathrm{nHz}$ GW data within the context of supermassive black hole binaries loosing energy in the form of gravitational waves~\cite{EPTA:2023xxk,NANOGrav:2023hfp,Huang:2023chx} as well within cosmological setups like cosmic strings~\cite{Ellis:2023tsl,Wang:2023len,Lazarides:2023ksx,Eichhorn:2023gat,Chowdhury:2023opo,Antusch:2023zjk,Yamada:2023thl,Ge:2023rce,Basilakos:2023xof}, domain walls~\cite{Kitajima:2023cek,Blasi:2023sej,Gouttenoire:2023ftk,Lu:2023mcz,Babichev:2023pbf,Gelmini:2023kvo,Guo:2023hyp,Zhang:2023nrs}, primordial first-order phase transitions~\cite{Fujikura:2023lkn,Franciolini:2023wjm,Bringmann:2023opz,Addazi:2023jvg,Bai:2023cqj,Han:2023olf,Zu:2023olm,Ghosh:2023aum,Xiao:2023dbb,Li:2023bxy,DiBari:2023upq,Cruz:2023lnq,Gouttenoire:2023bqy,Ahmadvand:2023lpp,An:2023jxf,Salvio:2023ynn}, inflation~\cite{Vagnozzi:2023lwo,Frosina:2023nxu,Liu:2023ymk,Unal:2023srk,Bari:2023rcw,Das:2023nmm,Jiang:2023gfe,Gorji:2023sil}, scalar-induced gravitational waves~\cite{Balaji:2023ehk,Bhaumik:2023wmw,Franciolini:2023pbf,Inomata:2023zup,Cai:2023dls,Wang:2023ost,Yi:2023mbm,Choudhury:2023fwk,HosseiniMansoori:2023mqh,Choudhury:2023fjs} as well as within other more exotic scenarios~\cite{Choudhury:2023kam,Wu:2023pbt,Gouttenoire:2023nzr,Inomata:2023drn,Datta:2023vbs}. The interested reader can study~\cite{Ellis:2023oxs} for a recent review on the different possible interpretations of the NANOGrav/PTA GW signal.

Within this work, we will focus on a non-singular bouncing cosmological scenario~\cite{Mukhanov:1991zn,Brandenberger:1993ef}
which actually postulates a Universe being in a contracting phase before transitioning at some point (bounce) to the expanding  Hot Big Bang (HBB) era. Interestingly enough, non-singular bouncing cosmology is free of the initial singularity problem present in inflationary cosmology~\cite{Borde:1996pt} being able to address as well the horizon and flatness problems~\cite{Turner:1994hv}. It can also account for a nearly scale-invariant primordial curvature power spectrum on large scales~\cite{Lilley:2015ksa, Battefeld:2014uga, 
Peter:2008qz}, being thus in excellent agreement with CMB observational data~\cite{Cai:2014bea,Cai:2014xxa}.

In order to realise a non-singular bouncing cosmological era,  one must postulate an effective violation of the null energy condition for a short time interval. This requirement can be naturally achieved within the context of extended theories of gravity~\cite{CANTATA:2021ktz,Nojiri:2006ri,Capozziello:2011et}. Among other modified gravity setups, one can realise non-singular bounces within higher order gravity theories~\cite{Biswas:2005qr,Nojiri:2013ru, Miranda:2022wvr}, $f(R)$~\cite{Bamba:2013fha,Nojiri:2014zqa}, $f(T)$ 
~\cite{Cai:2011tc} and $f(Q)$ gravity~\cite{Bajardi:2020fxh}, within
Ekpyrotic~\cite{Khoury:2001wf,Khoury:2001bz} and  Pre-Big-Bang~\cite{Veneziano:1991ek} scenarios, massive 
gravity~\cite{Cai:2012ag}, DHOST and cyclic~\cite{Lehners:2008vx,Banerjee:2016hom,Saridakis:2018fth} setups~\cite{Ilyas:2020qja,Ilyas:2020zcb,Zhu:2021whu}, as well with braneworld cosmology~\cite{Shtanov:2002mb,Saridakis:2007cf} and loop quantum gravity~\cite{Wilson-Ewing:2012lmx,Barca:2021qdn}.

In this work, we will study in a model-independent way the production of scalar-induced gravitational waves (SIGWs) and primordial black holes (PBHs) within non-singular bouncing cosmology~\footnote{We need to note here that there have been already some first works connecting PBHs with bouncing cosmology both at the analytical~\cite{Carr:2011hv,Quintin:2016qro,Chen:2016kjx,Clifton:2017hvg} as well as at the numerical level~\cite{Chen:2022usd}. PBH production was explored as well during the HBB  era but solely within the context of $f(R)$ theory of gravity~\cite{Banerjee:2022xft}.} in relation with the NANOGrav $\mathrm{nHz}$ GW data, based on our previous work~\cite{Papanikolaou:2024fzf}. In particular, we will focus on matter bouncing cosmological setups where one is inevitably met with a scale-invariant curvature power spectrum on CMB scales and a growth of the primoridal curvature perturbations on super-horizon scales during the matter contracting phase~\cite{Chen:2016kjx}, leading ultimately to an enhanced PBH formation on small scales and an abundant production of scalar-induced gravitational waves (SIGWs).

The paper is organised as follows: In \Sec{sec:bounce}, we study in a model-independent way the background and the perturbation evolution within non-singular matter bouncing cosmology, deducing ultimately the primordial curvature power spectrum at horizon crossing time during the HBB expanding era. Followingly, in \Sec{sec:PBH}, we investigate the production of PBHs and compute their relative contribution to dark matter within the framework of peak theory. Then, in \Sec{sec:SIGW}, we explore the GW background induced at second order in cosmological perturbation theory by scalar metric perturbations (SIGWs) during the HBB expanding era, in relation with the recently released NANOGrav $\mathrm{nHz}$ GW data checking also its detectability by future GW observatories such as LISA and ET. Finally, in \Sec{sec:Conclusions} we conclude.

\section{Non-singular bouncing cosmology}\label{sec:bounce}
We consider here a non-singular bouncing model which experiences initially a matter-dominated contracting  phase, passes then through a non-singular bounce, and transitioning ultimately finally into the HBB expanding phase, being radiation-dominated. The bouncing phase starts and end at $t_{-}$ and $t_{+}$ respectively with $t=0$ being the cosmic time when the Hubble parameter becomes zero, i.e. $H = 0$ (bouncing point).

Regarding the background dynamical evolution, one can show that under the assumptions mentioned above the scale factor can be phenomenologically written in every cosmic epoch as follows: ~\cite{Cai:2012va,Cai:2013kja}.

\textit{(i) Contracting Phase ($t<t_{-}$):}
\begin{equation}\label{eq:a_contracting}
 a(t)=a_{-} \left(\frac{t-\tilde{t}_{-}}{t_{-}-\tilde{t}_{-}}\right)^{2/3}~,
\end{equation}
with $a_{-}$ being the scale factor at the end of the matter contracting phase/beginning of the bouncing phase, i.e. at $t = t_{-}$. If $H_-$ stands for the Hubble parameter at $t_{-}$,  one then gets that $t_{-}-\tilde{t}_{-}=\frac{2}{3H_{-}}$. Note that $\tilde{t}_-$ in \Eq{eq:a_contracting} is a negative integration constant introduced here so as to achieve an analytic continuation of $H$ at $t_-$. 

\textit{(ii) Bouncing Phase ($t_-\leq t\leq t_+$):}
\begin{equation}
a(t)=a_\text b e^{\frac{\Upsilon t^2}{2}}~,
\label{abounce}
\end{equation}
with $a_\text b$ being the scale factor at $t=0$ and $\Upsilon$ a model parameter depending on the particulars of the underlying theory of gravity being responsible for the bounce. Imposing analytic continuation of $a$ at $t_-$ one finds that $a_{-}=a_\text b \exp [\Upsilon t_{-}^2/{2}]$, while the Hubble parameter during the bouncing phase reads as
\begin{equation}
\label{bouncingH}
 H(t)=\Upsilon t~.
\end{equation}

\textit{(iii) Hot Big Bang Expanding Phase ($t> t_+$):}
\begin{equation}\label{eq:a_HBB_era_cosmic_time}
 a(t)=a_{+} \left(\frac{t-\tilde{t}_{+}}{t_{+}-\tilde{t}_{+}}\right)^{1/2}~,
\end{equation}
with $t_{+}=H_{+}/\Upsilon$ and $t_{+}-\tilde{t}_{+}=\frac{1}{2H_{+}}$. Requiring again the scale factor to be continuous at the end of the bouncing phase/beginning of the HBB era, i.e. at $t=t_{+}$, one acquires $a_{+}=a_\text b e^{\frac{\Upsilon t_{+}^2}{2}}$. 

Going at the perturbative level, one should highlight that an initial sub-horizon mode $k$ crosses the cosmological horizon during the contracting phase, re-enters again in and around the bouncing phase and, after re-crossing the cosmological horizon gets again in causal contact with the observable Universe during the HBB phase. In the following sections, being agnostic on the underlying gravity theory driving the bounce, we study the dynamics of the perturbations during each of the three aforementioned cosmic eras in a model-independent way.

We  will work in terms of the Fourier mode of the Mukhanov-Sasaki (MS) variable $v_k$, being associated to the comoving curvature perturbation $\mathcal{R}_k$ as $v_k=z \mathcal R_k$ where $z=\frac{a \sqrt{\rho+p}}{c_s H\Mp}$. $c_\mathrm{s}$ denotes the curvature perturbation sound speed and $\Mp$ the reduced Planck mass, while $\rho$ and $p$ are the energy and pressure densities, correspondingly. Introducing the conformal time $\eta$ defined as $\mathrm{d}\eta \equiv \mathrm{d}t/a$, the evolution of $v_k$ will be governed by the following equation:
\begin{equation}
\label{EoM}
v_k''+\left(c_\mathrm{s}^2 k^2-\frac{z''}{z}\right)v_k=0~,
\end{equation}
Imposing thus vacuum Bunch-Davies initial conditions deep in the sub-horizon regime, i.e. $v_k(k\gg aH) \simeq \frac{e^{-ik\eta}}{\sqrt{2k}}$, one finds that $v_k$ during the matter contracting phase can be recast as
\begin{equation}
\label{eq:MS_solution_contracting}
v^\mathrm{m}_k=\frac{\sqrt{\pi(-\eta)}}{2} H_{3/2}^{(1)}[c_{s,m} k (-\eta)]~,
\end{equation}
where $H_{3/2}^{(1)}$ is the $\frac{3}{2}$-order Hankel function of the first kind. Thus, the reduced curvature power spectrum  $\mathcal{P}_\mathcal{R}(k)$, related to $\mathcal{R}(k)$ as $\mathcal{P}_\mathcal{R}(k)\equiv \frac{k^3}{2\pi^2}|\mathcal{R}_k|^2$, will read as
\beq\label{eq:P_R_contracting}
\mathcal{P}_\mathcal{R}(k) = \frac{k^3}{2\pi^2}\left|\frac{v_k}{z}\right|^2 = \frac{c^2_\mathrm{s,m}k^3(-\eta)}{24\pi\Mp^2a^2}\left|H_{3/2}^{(1)}[c_\mathrm{s,m} k (-\eta)]\right|^2.
\eeq
As one can see from \Eq{eq:P_R_contracting}, on large scales, i.e. $c_\mathrm{s,m}k\ll |aH|$, we get an almost scale invariant but dynamical $\mathcal{P}_\mathcal{R}(k)$, reading as $\mathcal{P}_\mathcal{R}(k) \simeq \frac{a^3_{-}H^2_{-}}{48\pi^2c_\mathrm{s,m}\Mp^2a^3}$, something which is in contrast with the conservation of the curvature perturbations on superhorizon scales in an expanding Universe. More specifically, within non-singular matter bouncing cosmology, one gets a curvature power spectrum $\mathcal{P}_\mathcal{R}(k)$ growing with time, since during the matter contracting phase $a$ is decreasing with time. On small scales,  i.e. $c_\mathrm{s,m}k\gg |aH|$, we obtain from \Eq{eq:P_R_contracting} that $\mathcal{P}_\mathcal{R}(k)\simeq \frac{a^3_{-}H^2_{-}}{12\pi^2c_\mathrm{s,m}\Mp^2a^3}\left(\frac{c_\mathrm{s,m}k}{aH}\right)^2 $.

\begin{figure*}[t!]
\begin{center}
\includegraphics[width=0.796\textwidth]{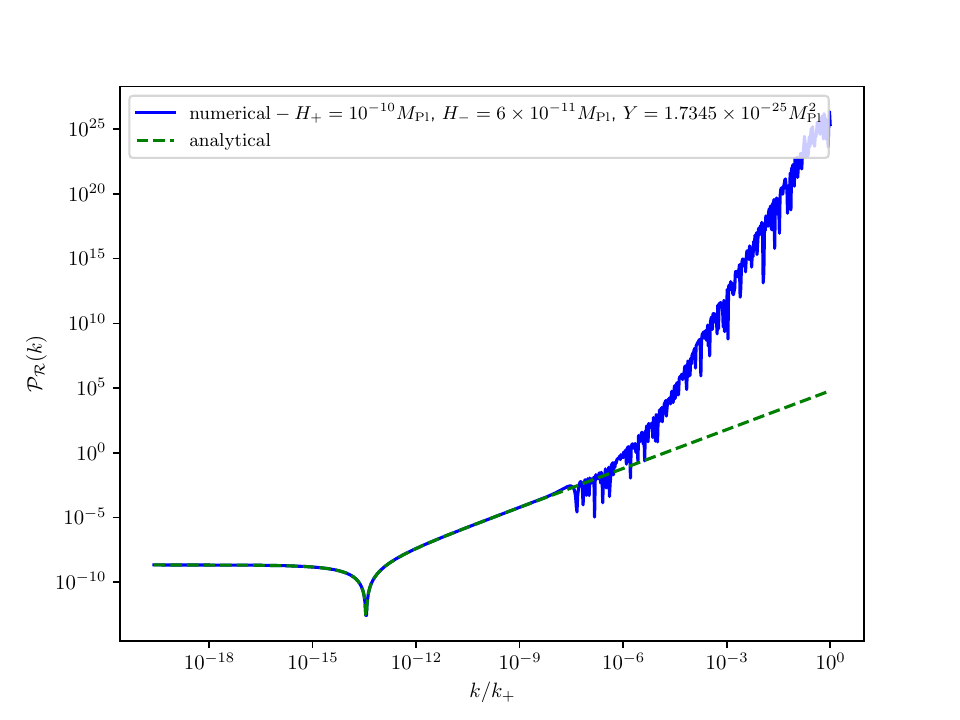}
\caption{{\it{The solid blue curve corresponds to the full curvature power spectrum, for $H_{+} = 10^{-10}\Mp$, $H_{-} = 6\times 10^{-11}\Mp$ and $\Upsilon = 1.7345\times 10^{-10}\Mp^2$. The dashed green curve corresponds to the analytic approximation for $\mathcal{P}_{\mathcal{R}}(k)$ up to linear order in $k$, while the dashed red curve  depicts the  analytic approximation for $\mathcal{P}_{\mathcal{R}}(k)$ up $\mathcal{O}(k^{9/2})$.}}}
\label{fig:P_R_full_vs_analytic}
\end{center}
\end{figure*}

Requiring at the end continuity of $v_k$ and $v'_k$ at $t=t_-$ and $t=t_+$, one can obtain in the limit of a short duration bouncing phase the MS variable $v_k$ during the HBB radiation-dominated phase, $v^\mathrm{RD}_k$. Finally, after a long but straightforward calculation [See~\cite{Papanikolaou:2024fzf} for more details] and accounting also for the conservation of the curvature perturbation on super-horizon scales during an expanding phase, the primordial curvature power spectrum $\mathcal{P}_\mathcal{R}(k)$ at horizon crossing time, i.e. $k=aH$, during the HBB expanding era will read as
\begin{eqnarray}\label{eq:P_R_anal}
\mathcal{P}_{\mathcal{R}}(k) &\simeq & \frac{0.7 \Upsilon^8 \cos^2{A}^2}{c_\mathrm{s,m}^3 H_-^4 H_+^2 \pi^2 (H_+^2 + 2 \Upsilon)^4} \nonumber \\ &&-\frac{1.4 B^2 \sqrt{c_\mathrm{s,m}} \Upsilon^{17/2}
     \cos{A} \sin{A} \sqrt{k}}{c_{s,m}^3 H_-^4 H_+^2 \pi^2 (H_+^2 + 2 \Upsilon)^4} \nonumber \\ && +\frac{\Upsilon^5 \biggl[0.7 c_\mathrm{s,m} H_-^3 \sin{A}^2 + 
     0.9 B^2
       \Upsilon \cos{A}\left(-\frac{2  \Upsilon^2  \cos{A}}{
         H_+ (H_+^2 + 2 \Upsilon)} + 
         \sqrt{\Upsilon} \sin{A}\right)\biggr] k}{4 c_\mathrm{s,m}^3 H_-^5 H_+^2 \pi^2 \left(1 + \frac{H_+^2}{
     2 \Upsilon}\right)^2 (H_+^2 + 2 \Upsilon)^2}, \nonumber \\  
\end{eqnarray}
where $A=(H_- + H_+)/\sqrt{\Upsilon}$ and $B=\sqrt{H_-/\Upsilon}$. 
From \Eq{eq:P_R_anal}, one can infer that the first term gives us a scale-invariant power spectrum on large scales, favored by CMB observational probes, while the second and the third terms lead to an enhancement of $\mathcal{P}_{\mathcal{R}}(k)$ on small scales, responsible for PBH formation and SIGW production. From \Fig{fig:P_R_full_vs_analytic}, it is shown clearly, that the analytic approximate expression \eqref{eq:P_R_anal} for $\mathcal{P}_{\mathcal{R}}(k)$ (green dashed curve) can reproduce quite well the full numerical result (blue curve) at least within the perturbative regime, i.e. where $\mathcal{P}_{\mathcal{R}}(k)<1$. As we start to probe the non-perturbative regime for high values of $k$, one needs to expand $\mathcal{P}_{\mathcal{R}}(k)$ to higher orders in $k$ so as to account for the full non-perturbative behavior. 

At this point,  we need to stress that the enhancement of curvature perturbations on small scales is a generic feature of any non-singular matter bouncing cosmological setup.
This is can be understood by the the growth of the curvature perturbations on super-horizon scales during the matter contracting phase, independently of the parametrisation of the scale factor during the bouncing phase [See \Eq{eq:P_R_contracting}]. Notably, when passing from a contracting to an expanding phase, both the amplitude and the shape of the curvature power spectrum remain unchanged due to a no-go theorem~\cite{Quintin:2015rta,Battarra:2014tga}, independently of the duration of the bouncing phase.
One then can get a generic non-``fine-tuned" mechanism for the growth of curvature perturbations on small scales, responsible for PBH formation and an abundant production of induced GWs.

\section{Primordial black hole formation in the expanding Hot Big Bang era}\label{sec:PBH}

Having obtained above an enhanced $\mathcal{P}_{\mathcal{R}}(k)$ on small scales, one is met inevitably with the formation of PBHs when the energy density contrast $\delta$ of a local overdensity region becomes greater than a critical threshold value $\delta_c$. Since PBHs constitute a viable non-particle candidate for dark matter, we can compute firstly the PBH abundance and the relevant contribution of PBHs to the present dark matter density. 

In order to perform such an analysis we need to relate first the  energy density contrast $\delta$ with the comoving curvature perturbation $\mathcal{R}$. In the linear regime, i.e. $\mathcal{R}\ll 1$, one can show that $\delta$ and $\mathcal{R}$ are related as
\beq\label{eq:zeta_vs_delta:linear}
\frac{\delta\rho}{\rho_\mathrm{b}}\simeq -\frac{1}{a^2H^2}\frac{2(1+w)}{5+3w}\nabla^2\mathcal{R}(r) \Longrightarrow \delta_k =  \frac{k^2}{a^2H^2}\frac{2(1+w)}{5+3w}\mathcal{R}_k,
\eeq
where one may notice the presence of a natural $k^2$ damping relevant for the large scales that cannot be observed.

However, since the process of formation of PBHs is a fully non-linear process one need to account for the full non-linear relation between  $\delta$ and $\mathcal{R}$. Interestingly enough, one can infer that the smoothed energy density contrast, $\delta_\mathrm{m}$, relates with the linear energy 
density contrast $\delta_l$, given by \Eq{eq:zeta_vs_delta:linear}, as~\cite{DeLuca:2019qsy,Young:2019yug}
\beq\label{eq:delta_m_smoothed}
\delta_\mathrm{m} = \delta_l - \frac{3}{8}\delta^2_l,
\eeq
where sub-horizon scales have been smoothed out so as to account for the cloud-in-cloud problem, while larger scales are suppressed due to the $k^2$ damping seen in \Eq{eq:zeta_vs_delta:linear}. The smoothed linear energy density contrast, $\delta_l$ can be defined as
\beq\label{eq:smoothed_delta_l}
\delta^R_l \equiv \int \mathrm{d}^3\vec{x}^\prime W(\vec{x},R)\delta(\vec{x}-\vec{x}^\prime),
\eeq
where $W(\vec{x},R)$ stands for a Gaussian window function, written in $k$ space as~\cite{Young:2014ana}
\beq\label{eq:Gaussian_window_function}
\tilde{W}(R,k) = e^{-k^2R^2/2},
\eeq
with $R$ denoting the smoothing scale, roughly equal to the comoving horizon scale $R=(aH)^{-1}$ for nearly monochromatic PBH mass distributions. From \Eq{eq:zeta_vs_delta:linear}, one can define as well the smoothed variance of $\delta$ as
\beq\label{eq:sigma}
\begin{split}
\sigma^2 & \equiv \langle \left(\delta^{R}_l\right)^2\rangle = \int_0^\infty\frac{\mathrm{d}k}{k}\mathcal{P}_{\delta_l}(k,R)  \\ & = \frac{4(1+w)^2}{(5+3w)^2}\int_0^\infty\frac{\mathrm{d}k}{k}(kR)^4 \tilde{W}^2(k,R) \mathcal{P}_\mathcal{R}(k),
\end{split}
\eeq
where $\mathcal{P}_{\delta_l}(k,R)$ and $\mathcal{P}_{\mathcal{R}}(k)$ denote correspondingly the reduced energy density and curvature power spectra. 

With regard to the PBH mass, the latter is of the order of the mass within the cosmological horizon when PBHs form, i.e. at horizon crossing time, denoted as  $M_\mathrm{H}$, while its full mass spectrum will obey to the following  critical collapse scaling law:~\cite{Niemeyer:1997mt,Niemeyer:1999ak,Musco:2008hv,Musco:2012au},  
\beq\label{eq:PBH_mass_scaling_law}
M_\mathrm{PBH} = M_\mathrm{H}\mathcal{K}(\delta-\delta_\mathrm{c})^\gamma.
\eeq
In the above expression, $\gamma$ is a critical exponent, depending on the equation-of-state (EoS) paremeter at the time of PBH formation. For radiation, $\gamma \simeq 0.36$. On the other hand, the parameter $\mathcal{K}$ depends both on the EoS parameter as well as on the shape of the collapsing overdensity region. In what follows, we will consiser a typical value for $\mathcal{K}\simeq 4$ based on numerical simulations of PBH formation during the radiation-dominated era~\cite{Musco:2008hv}. 

Concerning now the PBH formation threshold value, $\delta_\mathrm{c}$, it will depend, in general, on the shape of the collapsing curvature perturbation profile~\cite{Musco:2018rwt,Musco:2020jjb} as well as on the EoS parameter at the time of PBH formation~\cite{Harada:2013epa,Escriva:2020tak,Papanikolaou:2022cvo}. Since in our case, we consider spherical gravitational collapse of PBHs during the RD era, we will study only  the effect of the profile shape of the collapsing curvature power spectrum on $\delta_\mathrm{c}$, following closely the analysis described in in~\cite{Musco:2020jjb}. 

One then can straightforwardly show within the context of peak theory that the abundance $\Omega_\mathrm{PBH}(M)$, of PBHs with mass $M$, can be recast as
\beq\label{eq:beta_full_non_linear}
\Omega_\mathrm{PBH}(M) = \int_{\nu_\mathrm{c}}^{\frac{4}{3\sigma}}\mathrm{d}\nu\frac{\mathcal{K}}{4\pi^2}\left(\nu\sigma - \frac{3}{8}\nu^2\sigma^2 - \delta_{\mathrm{c}}\right)^\gamma \frac{\mu^3\nu^3}{\sigma^3}e^{-\nu^2/2},
\eeq
where $\nu_\mathrm{c} = \delta_{\mathrm{c},l}/\sigma$,  $\delta_{\mathrm{c},l}=\frac{4}{3}\left(1 -
\sqrt{\frac{2-3\delta_\mathrm{c}}{2}}\right)$ and $\mu$  stands for first moment of the
smoothed curvature power spectrum, being defined as
\beq
\begin{split}
\mu^2 & =\int_0^\infty\frac{\mathrm{d}k}{k}\mathcal{P}_{\delta_l}(k,R)\left(\frac{k}{aH}\right)^2 \\ & = \frac{4(1+w)^2}{(5+3w)^2}\int_0^\infty\frac{\mathrm{d}k}{k}(kR)^4 \tilde{W}^2(k,R)\mathcal{P}_\mathcal{R}(k)\left(\frac{k}{aH}\right)^2.
\end{split}
\eeq
Defining then, the contribution of PBHs to the dark matter abundance, $f_\mathrm{PBH}$, as
\beq\label{eq:f_PBH_def}
f_\mathrm{PBH}\equiv \frac{\Omega_\mathrm{PBH,0}}{\Omega_\mathrm{DM,0}},
\eeq
where the subscript $0$ refers to our present epoch, one finds straightforwardly that  $f_\mathrm{PBH}$ will read as
\beq\label{eq:f_PBH}
f_\mathrm{PBH} = \left(\frac{\beta(M)}{3.27 \times 10^{-8}}\right) \left(\frac{106.75}{g_{*,\mathrm{f}}}\right)^{1/4}\left(\frac{M}{M_\odot}\right)^{-1/2},
\eeq
where $M_\odot$ stands for the mass of the sun and $g_{*,\mathrm{f}}$ for the number of relativistic degrees of freedom, which for our case, considering PBHs in the very early Universe, is of the order of $100$~\cite{Kolb:1990vq}.

\begin{figure*}[t!]
\begin{center}
\includegraphics[width=0.796\textwidth]{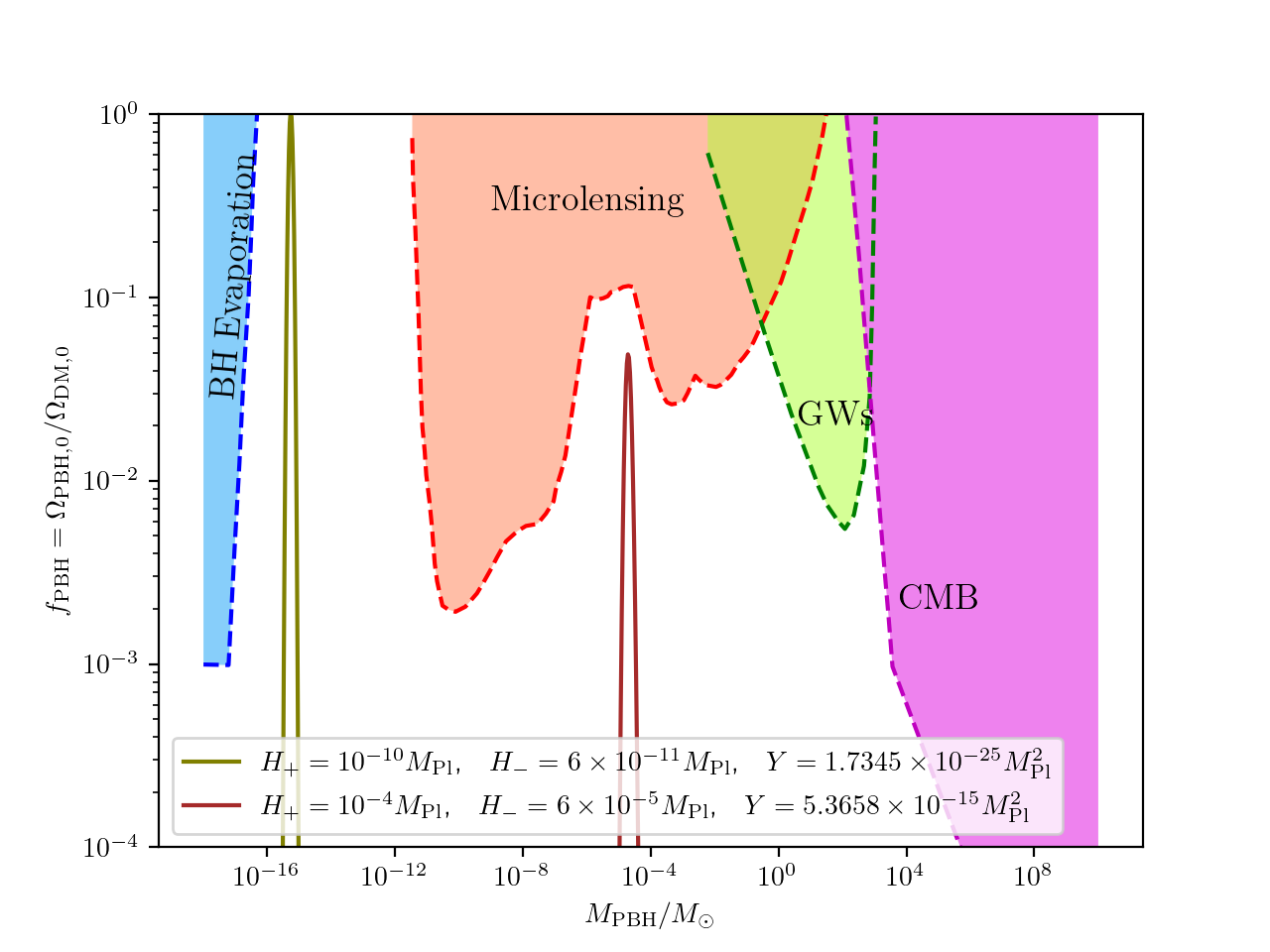}
\caption{{\it{The energy density contribution of PBHs to dark matter, expressed here as $f_\mathrm{PBH}=\Omega_\mathrm{PBH,0}/\Omega_\mathrm{DM,0}$ as a function of the mass of PBHs. The colored regions are excluded from evaporation (blue region), microlensing (red region), gravitational-wave (green region) and CMB (violet region) observational probes. The data for the constraints on $f_\mathrm{PBH}$ were obtained from~\cite{Green:2020jor}.}}}
\label{fig:f_PBH}
\end{center}
\end{figure*}

In \Fig{fig:f_PBH}, we depict the PBH energy density contribution to dark matter $f_\mathrm{PBH}$ as a function of the PBH mass for two different fiducial sets of the theoretical parameters involved, namely the Hubble parameter at the onset of HBB phase, $H_+$, the Hubble parameter at the end of the matter contracting phase $H_-$, and the bouncing parameter $\Upsilon$. We also superimpose observational constraints on $f_\mathrm{PBH}$ from evaporation (blue region)~\cite{Poulin:2016anj,Clark:2016nst,Boudaud:2018hqb,DeRocco:2019fjq,Laha:2019ssq}, microlensing (red region)~\cite{Macho:2000nvd,Niikura:2017zjd,Niikura:2019kqi,Zumalacarregui:2017qqd}, GW (green region)~\cite{Kavanagh:2018ggo,Chen:2019irf} and CMB (violet region)~\cite{Serpico:2020ehh} observational probes.

As one may infer from \Fig{fig:f_PBH}, we can produce PBH masses spanning many orders of magnitude, depending on the values of $H_+$, $H_-$ and $\Upsilon$. As we mentioned above, the PBH mass $M_\mathrm{PBH}$ will be of the order of the mass within the cosmological horizon at horizon crossing time during the HBB expanding era [See \Eq{eq:PBH_mass_scaling_law}]. One then obtains straightforwardly that $M_\mathrm{PBH}$ will scale with $H_+$, $H_-$, $\Upsilon$ and the comoving scale $k$ as 
\beq
M_\mathrm{PBH}\simeq M_\mathrm{H} = \frac{4\pi\Mp^2}{H_\mathrm{hc}(H_+,H_-,\Upsilon,k)} = \frac{\pi\Mp^2H_+\left(H^2_+ + 2\Upsilon\right)^2}{\Upsilon^2k^2},
\eeq
where $H_\mathrm{hc}$ is the Hubble parameter at horizon crossing time during the HBB era.

Notably, as one can see from \Fig{fig:f_PBH} for $H_+ = 10^{-4}\Mp$, $H_- = 6\times 10^{-5}\Mp$ and $\Upsilon = 5.3658 \times 10^{-15}\Mp^2$, we can easily produce PBH masses of the order of one solar mass, being the typical black hole progenitor masses probed by LIGO-VIRGO-KAGRA (LVK). On the othe hand, for $H_+ = 10^{-10}\Mp$, $H_- = 6\times 10^{-11}\Mp$ and $\Upsilon = 1.7345\times 10^{-25}\Mp^2$, one produces PBHs within the observationally unconstrained asteroid-mass window with $f_\mathrm{PBH}\simeq 1$, accounting thus for the totality of the dark matter

We need to highlight here that in order to stay within the linear regime, we impose a non-linear cut-off scale $k_\mathrm{NL}$ depending on $H_+$, $H_-$ and $\Upsilon$ such as that $\mathcal{P}_\mathcal{R}(k_\mathrm{NL}) = 0.1$. Considering scales smaller that the non-linear cut-off, i.e. $k>k_\mathrm{NL}$, one enters into the regime where cosmological perturbation theory is not valid anymore and no analytic treatment of our calculation can be adopted.

\section{Scalar induced gravitational waves}\label{sec:SIGW}

Let us now investigate in this section the generation of GWs induced by second order gravitational interactions by enhanced scalar/curvature perturbations~\cite{Matarrese:1992rp,Matarrese:1993zf,Matarrese:1997ay,Mollerach:2003nq}  [See~\cite{Domenech:2021ztg} for a review].
Working in the Newtonian gauge~\footnote{As proved in~\cite{Hwang:2017oxa,Tomikawa:2019tvi,DeLuca:2019ufz,Yuan:2019fwv,Inomata:2019yww,Domenech:2020xin,Chang:2020tji}, there is no gauge dependence of SIGWs produced during a RD era, like in our case, due to the exponential decay of the scalar perturbations, in the late-time limit.}, the perturbed Friedman-Lema\^itre-Robertson-Walker (FLRW) metric can  be recast as
\bea
\label{metric decomposition with tensor perturbations}
\mathrm{d}s^2 = a^2(\eta)\left\lbrace-(1+2\Phi)\mathrm{d}\eta^2  + \left[(1-2\Phi)\delta_{ij} + \frac{h_{ij}}{2}\right]\mathrm{d}x^i\mathrm{d}x^j\right\rbrace \, ,
\eea
where $\Phi$ stands for the first-order scalar perturbation while $h_{ij}$ denotes the second-order tensor perturbation. One then can straightforwardly derive the equation of motion in the Fourier space which will read as~\cite{Ananda:2006af,Baumann:2007zm}:
\beq
\label{Tensor Eq. of Motion}
h_\boldmathsymbol{k}^{s,\prime\prime} + 2\mathcal{H}h_\boldmathsymbol{k}^{s,\prime} + k^{2}h^s_\boldmathsymbol{k} = 4 S^s_\boldmathsymbol{k}\, ,
\eeq
where 
$s = (+), (\times)$ denotes the two GR tensor polarisation modes and $S^s_\boldmathsymbol{k}$ accounts for a source term written as~\cite{Kohri:2018awv,Espinosa:2018eve}
\beq
\label{Source}
S^s_\boldmathsymbol{k}  =
\int\frac{\mathrm{d}^3 q}{(2\pi)^{3/2}}e^{s}(\boldmathsymbol{k},\boldmathsymbol{q})F(\boldmathsymbol{q},|\boldmathsymbol{k-q}|,\eta)\phi_\boldmathsymbol{q}\phi_\boldmathsymbol{k-q},
\eeq
with $e^{s}(\boldmathsymbol{k},\boldmathsymbol{q})\equiv e^{s}_{ij}(\boldmathsymbol{k}) q^iq^j $ and $e^{(s)}_{ij}$  standing for two GR polarisation tensors defined as
\begin{eqnarray}
e^{(+)}_{ij}(\boldmathsymbol{k}) \equiv \frac{1}{\sqrt{2}}\left[e_i(\boldmathsymbol{k})e_j(\boldmathsymbol{k}) - \bar{e}_i(\boldmathsymbol{k})\bar{e}_j(\boldmathsymbol{k})\right], \\ 
e^{(\times)}_{ij}(\boldmathsymbol{k}) \equiv \frac{1}{\sqrt{2}}\left[e_i(\boldmathsymbol{k})\bar{e}_j(\boldmathsymbol{k}) + \bar{e}_i(\boldmathsymbol{k})e_j(\boldmathsymbol{k})\right],
\end{eqnarray}
where $e_i(\boldmathsymbol{k})$,  $\bar{e}_i(\boldmathsymbol{k})$ and $\boldmathsymbol{k}/k$ form an orthonormal basis. In \Eq{Source}, we express $\Phi$  as $\Phi_k(\eta) = T_\Phi(\tilde{x})\phi_k$ with $\tilde{x}=k\eta$, where $\phi_k$ denoting the value of $\Phi$ at some reference time $\tilde{x}_0$, here taking it to be the horizon crossing time, while $T_\Phi(\tilde{x})$ is a transfer function.

One then can solve \Eq{Tensor Eq. of Motion} with analytical methods, with its solution reading as~\cite{Kohri:2018awv}
\beq
\label{tensor mode function}
h^s_\boldmathsymbol{k} (\eta)  =\frac{4}{a(\eta)} \int^{\eta}_{\eta_\mathrm{d}}\mathrm{d}\bar{\eta}\,  G^s_\boldmathsymbol{k}(\eta,\bar{\eta})a(\bar{\eta})S^s_\boldmathsymbol{k}(\bar{\eta}),
\eeq
where $G^s_{\boldmathsymbol{k}}(\eta,\bar{\eta})$ stands for the Green function, derived from the homogeneous equation 
\beq
\label{Green function equation}
G_\boldmathsymbol{k}^{s,\prime\prime}(\eta,\bar{\eta})  + \left( k^{2} -\frac{a^{\prime\prime}}{a}\right)G^s_\boldmathsymbol{k}(\eta,\bar{\eta}) = \delta\left(\eta-\bar{\eta}\right),
\eeq
with the boundary conditions $\lim_{\eta\to \bar{\eta}}G^s_\boldmathsymbol{k}(\eta,\bar{\eta}) = 0$ and $ \lim_{\eta\to \bar{\eta}}G^{s,\prime}_\boldmathsymbol{k}(\eta,\bar{\eta})=1$.  

At the end, we can deduce the tensor power spectrum $\mathcal{P}_{h}(\eta,k)$ defined with the use of the equal-time two-point correlation function for the tensor perturbations as
\bea
\label{tesnor power spectrum definition}
\langle h^r_{\boldmathsymbol{k}_1}(\eta)h^{s,*}_{\boldmathsymbol{k}_2}(\eta)\rangle \equiv \delta^{(3)}(\boldmathsymbol{k}_1 - \boldmathsymbol{k}_2) \delta^{rs} \frac{2\pi^2}{k^3_1}\mathcal{P}^{(s)}_{h}(\eta,k_1),
\eea
where again $s=(\times)$ or $(+)$. Ultimately, the tensor power spectrum will be related with the GW spectral abundance defined as $\Omega_\mathrm{GW}(\eta,k)\equiv \frac{1}{\bar{\rho}_\mathrm{tot}}\frac{\mathrm{d}\rho_\mathrm{GW}(\eta,k)}{\mathrm{d}\ln k}$. On sub-horizon scales, where we use the flat spacetime approximation~\cite{Maggiore:1999vm}, one  can show that  $\Omega_\mathrm{GW}(\eta,k)$ will read as \beq\label{Omega_GW}
\Omega_\mathrm{GW}(\eta,k) = \frac{1}{24}\left(\frac{k}{\calH(\eta)}\right)^{2}\overline{\mathcal{P}^{(s)}_h(\eta,k)}.
\eeq

After a long but straightforward calculation, the tensor power spectrum $\mathcal{P}_{h}(\eta,k)$ will be recast as [see ~\cite{Kohri:2018awv,Espinosa:2018eve} for more details] 
\begin{equation}
\label{Tensor Power Spectrum}
\mathcal{P}^{(s)}_h(\eta,k) = 4\int_{0}^{\infty} \mathrm{d}v\int_{|1-v|}^{1+v}\mathrm{d}u   \left[ \frac{4v^2 - (1+v^2-u^2)^2}{4uv}\right]^{2} \times I^2(u,v,x)\mathcal{P}_\mathcal{R}(kv)\mathcal{P}_\mathcal{R}(ku)\,,
\end{equation}
where $I(u,v,x)$ denotes a kernel function, which contains information on the equation of state of the Universe during the era of GW production. In the case of a RD dominated epoch, like the HBB expanding phase, the GW spectral abundance $\Omega_\mathrm{GW}$ at horizon crossing time, will read as~\cite{Kohri:2018awv}
\beq\label{eq:Omega_GW_f}
\begin{split}
\Omega_\mathrm{GW}(\eta_\mathrm{hc},k)  & = \frac{1}{12}\int_{0}^{\infty} \mathrm{d}v\int_{|1-v|}^{1+v}\mathrm{d}u \left[ \frac{4v^2 - (1+v^2-u^2)^2}{4uv}\right]^{2}\\ & \times \mathcal{P}_\mathcal{R}(kv)\mathcal{P}_\mathcal{R}(ku)  \left[\frac{3(u^2+v^2-3)}{4u^3v^3}\right]^{2} \\ & \times \biggl\{\biggl[-4uv + (u^2+v^2-3)\ln \left| \frac{3 - (u+v)^{2}}{3-(u-v)^{2}}\right|\biggr]^2  \\ & + \pi^2(u^2+v^2-3)^2\Theta(v+u-\sqrt{3})\biggr\},
\end{split}
\eeq
where $\mathrm{hc}$ stands for horizon-crossing time during the HBB era.

Finally, accounting for entropy conservation between horizon-crossing time and the present epoch, one obtains that
\beq\label{Omega_GW_RD_0}
\Omega_\mathrm{GW}(\eta_0,k) = \Omega^{(0)}_r\frac{g_{*\mathrm{\rho},\mathrm{f}}}{g_{*\mathrm{\rho},0}}\left(\frac{g_{*\mathrm{S},\mathrm{0}}}{g_{*\mathrm{S},\mathrm{f}}}\right)^{4/3}\OmegaGW(\eta_\mathrm{f},k),
\eeq
where the subscript $0$ denotes the present era while $g_{*\mathrm{\rho}}$ and $g_{*\mathrm{S}}$ stand for the energy and entropy relativistic degrees of freedom. For our numerical purposes,  we take $\Omega_\mathrm{rad,0}\simeq 10^{-4}$~\cite{Planck:2018vyg}, $g_{*\mathrm{\rho},0}\simeq g_{*\mathrm{S},0}= 3.36$, $g_{*\mathrm{\rho},\mathrm{f}}\simeq g_{*\mathrm{S},\mathrm{f}} = 106.75$~\cite{Kolb:1990vq}.

\begin{figure*}[h!]
\begin{center}
\includegraphics[width=0.796\textwidth]{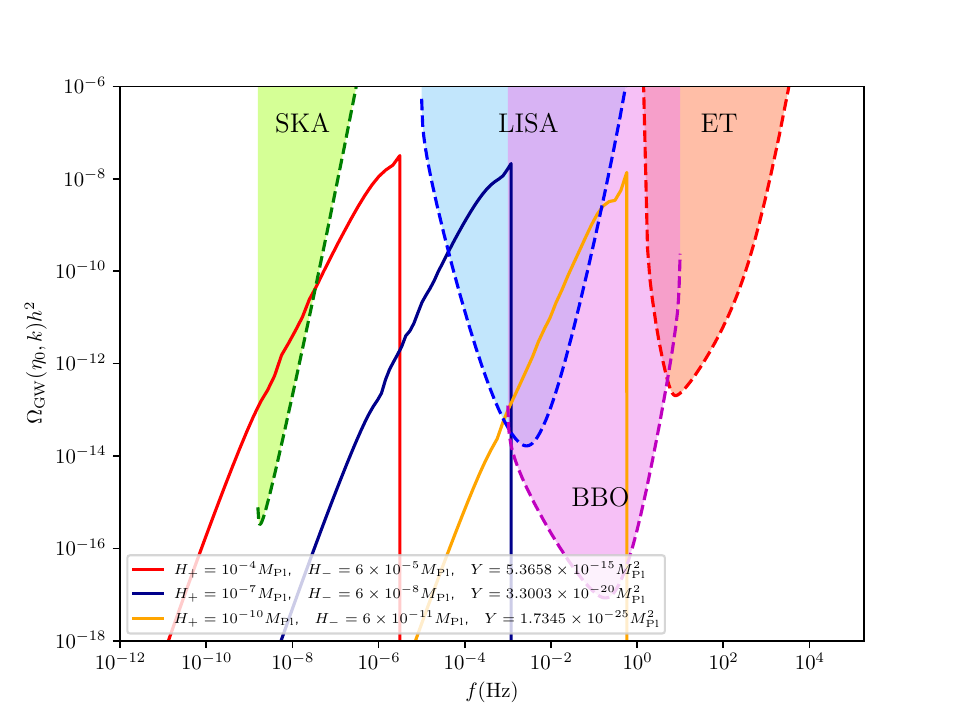}
\caption{{\it{The SIGW signal within non-singular matter bouncing cosmology for different values of the parameters $H_{+}$, $H_{-}$ and $\Upsilon$. We superimpose as well the frequency detection sensitivity bands of SKA~\cite{Janssen:2014dka}, LISA~\cite{LISACosmologyWorkingGroup:2022jok}, BBO~\cite{Harry:2006fi} and ET~\cite{Maggiore:2019uih} GW probes.}}}
\label{fig:Omega_GW}
\end{center}
\end{figure*}

In \Fig{fig:Omega_GW},  we show the current SIGW signal within non-singular matter bouncing cosmology as a function of the frequency $f$ defined as $f\equiv \frac{k}{2\pi a_0}$, for different sets of our theoretical parameters involved, namely $H_+$, $H_-$, and $\Upsilon$. Furthermore, we superimpose  the GW sensitivity bands of the future GW observational missions, namely the  Square Kilometer Arrays (SKA)~\cite{Janssen:2014dka}, the Einstein Telescope (ET)~\cite{Maggiore:2019uih}, the Big Bang Observer (BBO)~\cite{Harry:2006fi} and the Laser Inferometer Space Antenna (LISA)~\cite{LISACosmologyWorkingGroup:2022jok}.

As one may infer from \Fig{fig:Omega_GW}, at first $\Omega_\mathrm{GW}\propto f^2$ and then it decreases abruptly at an ultra-violet (UV) cut-off frequency $f_\mathrm{UV} \sim  \frac{k_\mathrm{NL}}{2\pi a_0}$, associated to the non-linear cutoff introduced in \Sec{sec:bounce} where $\mathcal{P}_\mathcal{R}(k_\mathrm{NL}) = 0.1$. Beyond this UV frequency perturbation theory breaks down making us loosing the analytical treatment of the present calculation. At these small scales (high frequencies), one is encountered with early structure formation leading to an abundant induced GW production. However, to probe such high-frequency induced GWs beyond the non-linear cut-off scale, it is necessary the running of fully numerical simulations~\cite{Fernandez:2023ddy,Dalianis:2024kjr}, something going beyond the scope of the present research work. Regarding now the scaling $f^2$ of the GW signal in the infrared (IR) frequency range, it is expected since as one can see from \Eq{eq:Omega_GW_f}, $\Omega_\mathrm{GW}\propto \mathcal{P}^2_\mathcal{R}$. Therefore, since $\mathcal{P}_\mathcal{R}\propto k$ (See \Eq{eq:P_R_anal}), one gets $\Omega_\mathrm{GW}\propto k^2 \propto f^2$.

\begin{figure*}[h!]
\begin{center}
\includegraphics[width=0.796\textwidth]{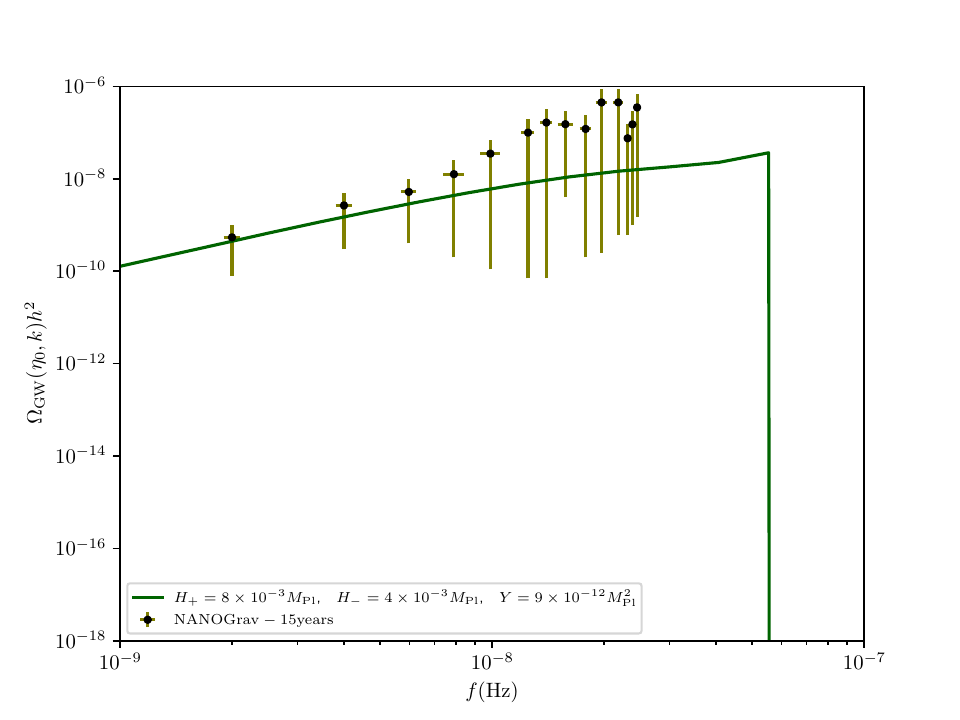}
\caption{{\it{ The SIGW signal within non-singular matter bouncing cosmology for $H_+=8\times 10^{-3}\Mp$, $H_-=4\times 10^{-3}\Mp$ and $\Upsilon = 9\times 10^{-12}\Mp^2$, in comparison with the NANOGrav GW data  \cite{NANOGrav:2023gor}.}}}
\label{fig:Omega_GW_NANOGrav}
\end{center}
\end{figure*}

In \Fig{fig:Omega_GW_NANOGrav}  we show as well the SIGW signal for $H_+ = 8\times 10^{-3}\Mp$, $H_- = 4\times 10^{-3}\Mp$ and $\Upsilon = 9 \times 10^{-12}\Mp^2$, superimposed with the recently Pulsar Time Array (PTA) GW data released by NANOGrav  \cite{NANOGrav:2023gor}. As one may see, our GW prediction for the fiducial values of $H_+$, $H_-$ and $\Upsilon$ reported above, peaks at $\mathrm{nHz}$ and it can explain quite well the PTA GW data. Remarkably, the frequency slope $+2$ predicted within our non-singular matter bouncing scenario, is well within the error-bars of the frequency slope reported by the NANOGrav collaboration~\cite{NANOGrav:2023gor}, i.e. $n_\mathrm{T} = 2.08\pm 0.3$ for a power-low GW fitting formula of the form $\Omega_\mathrm{GW}\propto \left(\frac{k}{k_0}\right)^{n_\mathrm{T}}$. 

Induced GWs produced within non-singular matter bouncing cosmology can then serve as one of the possible new physics explanations for the NANOGrav/PTA GW signal. A more rigorous likelihood analysis needs to be performed however in order to deduce the $H_+$, $H_-$ and $\Upsilon$ values best fitting the NANOGrav/PTA GW data, being beyond the scope of the present study.

\section{Conclusions}\label{sec:Conclusions}

The non-singular bouncing scenario constitutes an attractive early Universe cosmological paradigm. It is free of the initial singularity problem being capable as well to account for the flatness and horizon problems of HBB cosmology. Moreover, it is consistent with the large scale structure observational data, giving rise to a scale-invariant curvature power spectrum on large scales probed by CMB experiments. 

Notably, in this work we found that the no-singular matter bounce cosmological scenario can naturally give rise to enhanced curvature perturbations on small scales crossing the cosmological horizon during the HBB expanding phase. This is mainly due to the fact that curvature perturbations grow on super-horizon scales during the matter contracting phase freezing then in the subsequent bouncing and HBB expanding epochs. This result was found to be quite generic independently of the underlying gravity theory driving the bouncing phase.

Interestingly enough, these enhanced cosmological perturbations can collapse to form PBHs. Depending on the values of the theoretical parameters involved namely the Hubble parameter at the onset of HBB phase, $H_+$, the Hubble parameter at the end of the matter contracting phase $H_-$, and the bouncing parameter $\Upsilon$, one can give rise to PBHs in different mass ranges, in particular within the observationally unconstrained asteroid-mass range, where PBHs can account up to the totality of the dark matter.

Remarkably, these enhanced cosmological perturbations can induce as well gravitational waves (GWs) at second order in cosmological perturbation theory with a universal infrared (IR) frequency scaling of $f^2$, in excellent agreement with the recently released $\mathrm{nHz}$ GW data by NANOGrav/PTA collaboration. This SIGW signal can be potentially detectable as well by other GW observatories such as LISA and ET, depending on the values of the bouncing cosmological parameters at hand, acting thus as a novel portal giving us access to the conditions prevailing the Universe at its very first moments.

\begin{acknowledgments}
TP acknowledges the support of the INFN Sezione di Napoli \textit{initiativa specifica} QGSKY. TP  acknowledges as well the  contribution of the LISA Cosmology Working Group and the COST Action CA21136 ``Addressing observational tensions in cosmology with systematics and  fundamental physics (CosmoVerse)''.
\end{acknowledgments}


\bibliographystyle{JHEP} 
\bibliography{PBH}
\end{document}